\documentclass[12pt]{spieman}  
\usepackage{amsmath,amsfonts,amssymb}
\usepackage{graphicx}
\usepackage{setspace}
\usepackage{tocloft}
\usepackage{subcaption}
\usepackage{lineno}

\title{Characterization of Random Telegraph Noise in an H2RG X-ray Hybrid CMOS Detector}

\author[a]{William A. Bevidas Jr.}
\author[a,b]{Joseph M. Colosimo}
\author[a]{Abraham D. Falcone}
\author[a]{Timothy Emeigh}
\author[a]{Lukas R. Stone}
\author[a]{Kadri M. Nizam}
\author[a]{Brynn Bortree}
\author[a,c]{Jacob C. Buffington}
\author[a]{David N. Burrows}
\author[a]{Zachary E. Catlin}
\author[a]{Killian M. Gremling}
\author[a]{Md. Arman Hossen}
\author[a]{Collin Reichard}
\author[a]{Ana C. Scigliani}
\author[a,d]{Anthony J. Tavana}
\author[a]{Mitchell Wages}

\affil[a]{The Pennsylvania State University, Department of Astronomy and Astrophysics, 525 Davey Lab, University Park, USA, 16802}
\affil[b]{NASA Goddard Space Flight Center, X-ray Astrophysics Laboratory, Greenbelt, MD 20771}
\affil[c]{Montana State University, Department of Physics, 212 Montana Hall, Bozeman, MT 59717}
\affil[d]{Rochester Institute of Technology, Center for Detectors, 74 Lomb Memorial Drive, Rochester, NY 14623}

\cftpagenumbersoff{figure}
\cftpagenumbersoff{table} 
\begin{document} 
\maketitle

\begin{abstract}
Hybrid CMOS detectors (HCDs) have several excellent features as high-performance X-ray detectors, including rapid readout, deep-depletion silicon for high quantum efficiency, radiation hardness, and low power. Random telegraph noise (RTN) is a type of noise that can reduce the performance of HCDs and other CMOS sensors. After finding and quantifying RTN in the recently developed engineering grade Speedster-EXD550 HCDs, this form of noise has also been found in other X-ray HCDs. This paper aims to investigate its presence and characteristics in the relatively mature H2RG X-ray HCD and to compare it with that of the Speedster-EXD550. We use archival data taken with an H2RG X-ray HCD at two different temperatures to determine the percentage of pixels that are being impacted by RTN. We identify RTN in 0.42\% of pixels when the detector is operated at 140 K, while we are only able to identify RTN in 0.060\% of pixels when the detector is operated at 160 K. We characterize RTN in two Speedster-EXD550 detectors, identifying 5.1\% of pixels with RTN in one detector and 7.1\% in another, which is significantly more than the H2RG. These results verify the difference between two different HCDs and provide techniques that can be applied to future hybrid CMOS detectors. 

\end{abstract}



\begin{spacing}{1}   

\section{Introduction}
\label{sect:intro}

Modern charge-coupled devices (CCDs) are designed to achieve minimal noise and provide an accurate detection of photons, including individual photon detection and energy estimation in the X-ray energy band.\cite{janesick1985ccd} X-ray hybrid complementary metal-oxide semiconductor (CMOS) detectors (HCDs) are a newer type of photon detector composed of two separate silicon substrates that have been bump-bonded (``hybridized") together. The top, X-ray absorbing, silicon layer is designed to optimize quantum efficiency (QE) through the use of high-resistivity silicon, allowing for the bulk to be fully depleted. The bottom substrate, (\textit{i.e.}, the readout integrated circuit (ROIC)), is optimized for fast and efficient readout schemes. \cite{bongiorno2015measuring} The absorbing layer allows each X-ray photon to interact with the silicon, causing a charge cloud to pass through the substrate.
The charge, collected by the ROIC, is then amplified and the signal from each pixel is digitized using an analog-to-digital converter (ADC) to provide a digital number (DN) value that can be stored and analyzed. While these active pixel sensors offer faster readout rates, lower power consumption, and enhanced radiation hardness, reducing noise is still the most active area of research.\cite{falcone2007hybrid,falcone2021x} 

\subsection{The H2RG X-ray HCD}

\begin{figure}[t]
    \centering
    \includegraphics[width=44\columnsep]{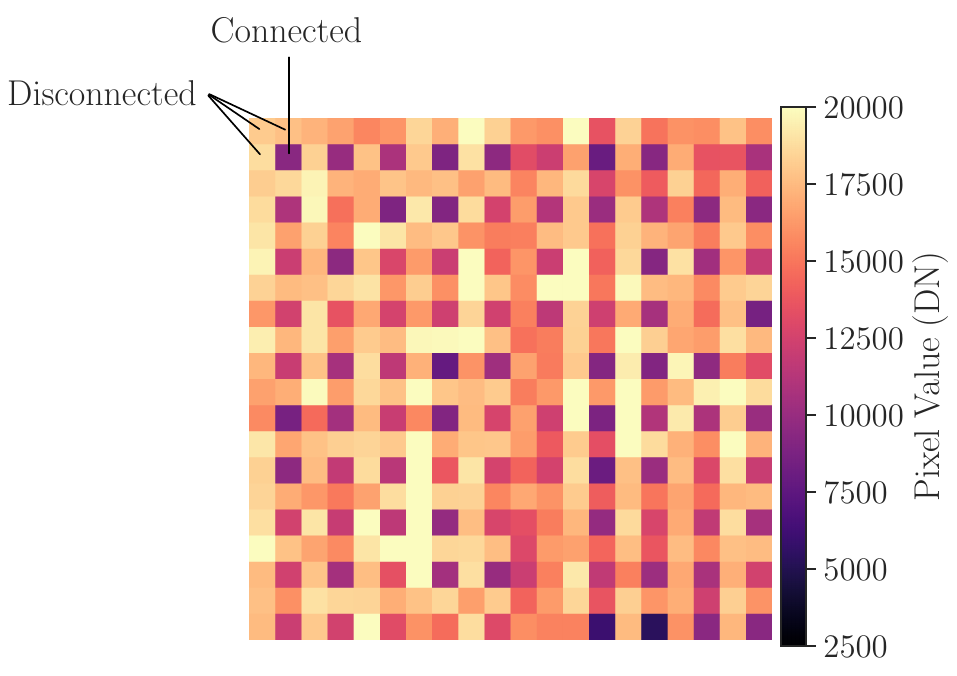}
    \vspace{0.3cm}
    \caption{A 20 $\times$ 20 segment of a raw image taken with the H2RG-122 at 140 K. The top left corner shows three pixels disconnected from the substrate and one connected to the substrate. This pattern repeats across the entirety of the detector.}
    \label{fig:h2}
\end{figure}

The HAWAII-xRG (HxRG) HCDs, developed by Teledyne Imaging Sensors (TIS), are used for detecting photons over a wide range of energies.\cite{blank2011hxrg,bai2008teledyne} In the HxRG line of detectors, the $x$ signifies the size of the ROIC array in multiples of 1024 $\times$ 1024 pixels. Hybrid Visible Silicon Imager (HyViSI) variations of HxRGs provide high sensitivity over the soft X-ray bandpass. The X-ray H2RG used in this investigation has a substrate with a 1024 $\times$ 1024 pixel array of 36-$\mu$m pixel pitch bonded to a ROIC with a 2048 $\times$ 2048 pixel array of 18-$\mu$m pixel pitch. One out of every four ROIC pixels is connected to the corresponding substrate pixel (see Figure \ref{fig:h2}). 

This specific X-ray H2RG design is meant to mitigate an issue of the H2RG's predecessors, the X-ray H1RGs, which suffered from interpixel capacitance crosstalk (IPC), where the charge being integrated directly onto the detector node caused a voltage change that allowed for signal to couple with neighboring pixels.\cite{bongiorno2009measurements} The increase in pixel size on the substrate reduces the IPC due to the increased spacing between pixels. \cite{prieskorn2013characterization} HxRGs use source followers (SFs) as amplifiers for each pixel because of their simple design and low noise characteristics.\cite{beletic2008teledyne} The SF uses a basic metal--oxide--semiconductor field-effect transistor (MOSFET) with a constant current supply; the voltage differential across the transistor gives a measurement of the charge integrated onto the sense node.

The H2RG is read out using a process called up-the-ramp integration. This process begins by resetting the detector to acquire a data frame with effectively zero integration time. Charge is then integrated onto the detector over a short period of time and read out. After this, more charge is integrated onto the detector without resetting the charge, allowing it to accumulate throughout the ramp, which typically consists of around 100 and 300 samples. The charge is reset at the end of the ramp and this process repeats. In some detectors, correlated double sampling (CDS) is used to correct for reset noise through in-pixel sampling and correction for the variable reset level. The H2RG does not have in-pixel CDS circuitry; however, reset noise is not a contributing factor because up-the-ramp integration does not involve resetting the detector. To find the charge integrated between exposures and correct for fixed-pattern noise, a pseudo-CDS process is used in which subsequent frames from the integration process are digitally subtracted during analysis. 

\subsection{Speedster-EXD550}

The Speedster-EXD550 is another type of HCD that has a substrate with a 550 $\times$ 550 pixel array of 40-$\mu$m pixel pitch bonded to a ROIC with a 550 $\times$ 550 pixel array of the same pitch.\cite{griffith2016speedster} Capacitive transimpedance amplifiers (CTIAs) are used in the Speedster-EXD550 detectors. In this pixel design, the voltage at the sense node is kept constant (unlike in a source-follower design), eliminating IPC in these detectors. This detector can operate in an event-driven readout mode that uses a comparator in each pixel to identify pixels with signals above the background noise for readout and further processing. The SpeedsterEXD-550 uses in-pixel CDS circuitry to correct for reset noise. 

\subsection{Noise}

Amplifier noise is the uncertainty in the charge measurement from the amplifier. The MOSFET in the amplifier contains three fundamental transistor noises: Johnson noise, flicker (1/f) noise, and random telegraph noise (RTN). Johnson noise arises from thermal noise within the MOSFET and is a type of frequency independent noise that can impact pixel output variability. \cite{johnson1928thermal} Flicker noise has a power spectral density which falls off with a frequency of 1/f. This type of noise is thought to result from defects in the silicon-oxide interface resulting in charge carriers in the MOSFET becoming mobile and/or dense. \cite{simoen1999flicker} The read noise floor for CCDs is known to be typically limited by 1/f noise, while the read noise floor is limited by RTN in CMOS sensors.\cite{janesick2006fundamental} Both noises are physically connected through the capture and emission of carriers in traps located within the amplifier. However, RTN typically involves single traps and carriers, while 1/f noise involves the combination of many RTN sources. An amplifier will be RTN or 1/f limited based on its size. A CCD amplifier is larger, so they will statistically contain more traps and will therefore be 1/f limited. Amplifiers in CMOS sensors are typically much smaller, so single RTN events in each pixel can cause deviations in the output and potentially make RTN a larger relative contribution to noise in CMOS detectors.

\subsection{RTN}

This paper will focus on a type of RTN, also known as MOSFET RTN, that is caused by an output change at or near the Si-SiO$_2$ gate interface of the amplifier. \cite{kohley2016random,goiffon2011evidence}. This change occurs when charge gets trapped within the amplifier and produces a potential barrier that causes a change in the output of the amplifier and, once released, a return to its original value. \cite{janesick2006fundamental} These charge traps cause a variation between two states, a high-output and a low-output state. \cite{bacon2005burst} Another type of RTN, causing a discrete change in a pixel’s dark current, has been observed in CMOS detectors but will not be focused on in this paper.\cite{bogaerts2002random,simoen2016random}

\begin{figure}
    \centering
    \includegraphics[width=1\linewidth]{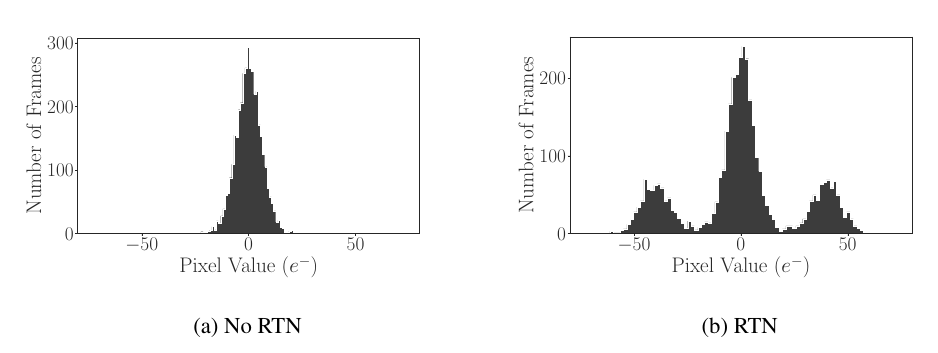}
    \vspace{0.1cm}
    \caption{Two CDS-processed pixels from the H2RG at 140 K using $\sim$3800 frames, where the triple peak pattern caused by RTN in (b) can be seen. Pixel values in (b) will give a much higher standard deviation than pixel values in (a), which in turn could significantly affect the total read noise of the detector, if enough RTN-affected pixels are present.}
    \label{fig:image2}
\end{figure}

When data are collected from a detector with CDS or analyzed using a psueo-CDS technique (subtracting one frame’s values from the subsequent frame) the signature of the RTN state gets modified. The process will either subtract two values in the same state or two values that are in different states, which causes the subtracted pixel values to fall within three potential distributions. When consecutive frames are in the same state (e.g. with no RTN event), the values follow a Gaussian distribution around zero (see Figure \ref{fig:image2}a). When the latter is true (i.e. different states from frame to frame), the values follow two Gaussian distributions, equidistant from the mean of the central distribution in both directions, with similar heights. The combination of these states forms a histogram with three distinct peaks (see Figure \ref{fig:image2}b).

\begin{figure}
    \centering
    \includegraphics[width=1\linewidth]{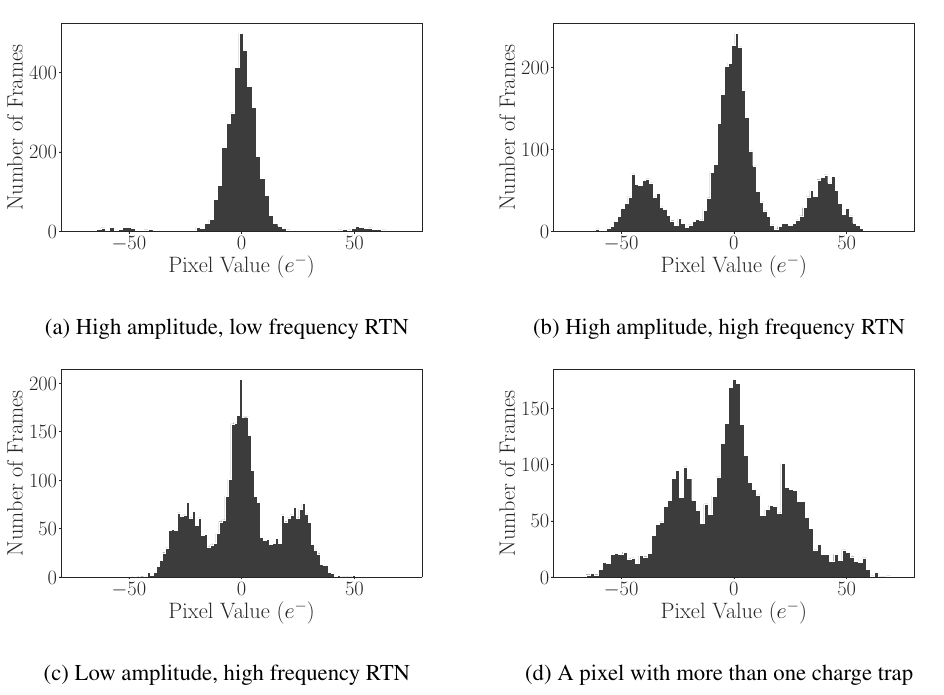}
    \vspace{0.1cm}
    \caption{Four pixels with different types of RTN that can occur in an HCD.}
    \label{fig:diffrtn}
\end{figure}

Each pixel of the detector can be represented by a histogram of the pixel's signal over many frames to observe if RTN affects that pixel by observing these peaks. All histograms that are influenced by RTN have unique characteristics that define its shape. The magnitude shift in the pixel value caused by a capture or release of an electron in a charge trap is defined as the \textit{amplitude}. This can be seen on the histogram of a CDS-subtracted pixel as the separation between the primary noise peak and the outer RTN peaks. For example, Figure \ref{fig:image2}b shows a pixel with clear RTN with a mean amplitude shift of approximately 40 e$^{-}$. The fraction of time that the pixel value is impacted by a change in state from the RTN trap is denoted as the \textit{frequency}. This is seen on the histogram as the occupancy of each RTN peak as a fraction of the primary noise peak. Figures \ref{fig:diffrtn}a and \ref{fig:diffrtn}b, respectively, demonstrate rare (low frequency) and recurrent (high frequency) state changes with major deviations in pixel value (high amplitude) in a charge trap. A trap that causes low amplitude and high frequency states can be seen in Figure \ref{fig:diffrtn}c. Multiple traps can be present, and a pixel can have more than one amplitude and frequency, shown in Figure \ref{fig:diffrtn}d. These are just a subset of examples that show how RTN manifests itself and how diverse the RTN population can be. 

The detection of RTN is important for X-ray astronomy, as this shift in the Gaussian noise can appear identical to an X-ray event. When an X-ray event occurs in a pixel with RTN, the inferred X-ray energy can be affected, resulting in inaccurate calculations of energy resolution, gain, pixel noise, and frame noise. Penn State, in collaboration with TIS, has been working to develop HCDs, including the Speedster-EXD550 \cite{colosimo2023initial}, as well as small-pixel detectors in different form factors including early engineering small-pixel128 detectors \cite{hull2018small} and the more recently developed small-pixel1024 detectors \cite{stone2024status}. Recent analysis confirms that both exhibit some RTN behavior, with the Speedster-EXD550 appearing to have significantly more RTN noise than was previously noted in other HCD X-ray detectors. This paper aims to assess the percentage and location of RTN-affected pixels in the H2RG, from which the developed methods can be used in these other detectors to find and remove events from RTN-affected pixels. For the rest of the paper, we will describe the data and methods used for detecting RTN. The results will consist of the percentage of RTN-affected pixels, characterization of the RTN in terms of amplitude and frequency, and how the amount of RTN in the H2RG compares with the Speedster-EXD550. 

\section{Methods}

This paper uses data collected from an H2RG with the serial number H2RG-122, produced by TIS. We utilize these H2RG data, which were previously utilized for quantum efficiency measurements (see Reference \citenum{colosimo2022measuring} for more details), that were taken between December 2020 and May 2021 at a range of different temperatures. A vacuum chamber facilitated data collection in low pressure environments, and liquid nitrogen was used to cool down the detectors. A cryo SIDECAR ASIC \cite{loose2005sidecar} and SIDECAR Acquisition Module developed by TIS were used as support electronics to operate the detector in the chamber. This section will explain how data were selected and what analyses were used for RTN detection.  

\subsection{Data}

Variable high noise states that are caused by RTN can be confused with X-ray events, so dark frames (i.e., data without the presence of an external X-ray source) are used for analysis. Thermal energy increases the detector's noise floor, so temperature-dependent analysis is needed for a consistent read noise. We are ultimately limited in our detection of low-frequency RTN by the amount of exposures that are taken. This means we need to have a significant amount of frames in order to maximize the probability of detecting RTN, particularly for pixels that only exhibit RTN occasionally (i.e. with low frequency). Since we are working with archival data, the temperature that includes the largest amount of exposures needs to be used for analysis. RTN that has low frequency and an amplitude lower than approximately 3 standard deviations of the read noise is also hard to find since the read noise will encapsulate the peak, so a lower read noise dataset should be used for analysis (i.e., a lower temperature). Data with the most dark frames and lowest read noise were chosen for analysis. 

We found 6,000 frames at 140 K and 9,400 at 160 K. The data taken at the lower temperature have a lower read noise; however, we have more frames at the higher temperature, so these were also analyzed, thus allowing us to increase the sample size and to observe the differences between the two temperatures. This set of data was processed with a subtraction algorithm that performs pseudo-CDS, boxcar smoothing, and removal of all pixels from the ROIC that are not connected to the substrate. The boxcar smoothing process occurs per frame and subtracts robust means of 15-pixel windows to a corresponding pixel in that row, removing row-wise local noise. We find that certain ramps have consistently higher read noise throughout all pixels. The reason for these high read noise states is not definitively understood. However, since these archival data were originally acquired for completely different measurements in the past and we are re-purposing them for the work reported here, it is possible that relative voltage and timing bias settings were not always at their optimized values for some of these data runs. To have a complete dataset with a consistent read noise, these deviant frames were removed from analysis. The final dataset consists of 3,766 frames taken at 140 K and 9,214 frames taken at 160 K. 

We measure the read noise by calculating the root-mean-square for every pixel in each of the ramps, so each ramp has a corresponding read noise measurement for all pixels. First, the median read noise is found for each pixel. Then, the mean and standard deviation of those values were found across the entirety of the detector array. Since a median is used for calculations, RTN has a negligible effect on the total read noise value. The read noise at both temperatures are 6.3 $\pm$ 0.6 e$^{-}$ and 13.5 $\pm$ 1.0 e$^{-}$, respectively. Neither the dark current nor read noise should have such a large change in a direct fashion. Therefore, some of the noise that is being called read noise is additional noise that is not dark current. This could be lower amplitude/frequency RTN that we cannot detect or characterize. Two Speedster-EXD550 detectors, FPM 23056 and FPM 23057, are chosen for analysis out of eight total detectors that have a wide range of RTN characteristics. \cite{colosimo2024characterization} Both datasets include 1,943,000 frames taken at 233 K. The read noise for both detectors at this temperature are 19.5 $\pm$ 0.3 e$^{-}$ and 19.0 $\pm$ 0.4 e$^{-}$, respectively.

\subsection{Methods}

The RTN-detection algorithm we employed uses histograms from each pixel to conduct the analysis. Specific thresholds are chosen to remove other types of noise that may affect a detector, so analysis can focus on finding RTN. The histograms are processed through the analysis pipeline to determine the presence of RTN by finding pixels with multiple peaks and analyzing them for RTN signatures. Once these are found, an amplitude and frequency can be calculated. Simulations were then created to determine the false-positive rate of the algorithm, the thresholds, and how different amplitudes and frequencies affect the detection rate.

\subsubsection{RTN Detection Algorithm} 

\begin{figure}
    \centering
    \includegraphics[width=\linewidth]{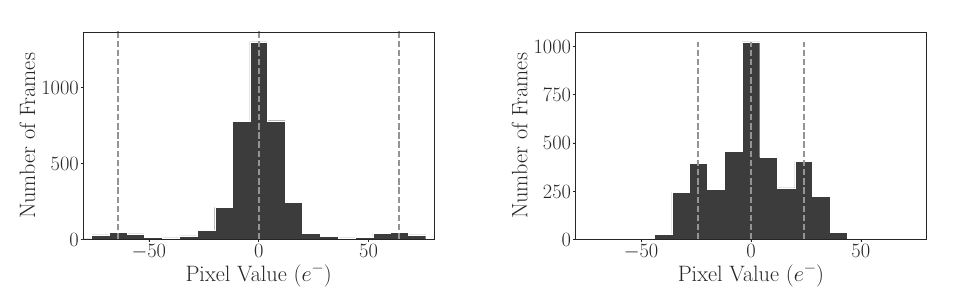}
    \vspace{0.1cm}
    \caption{Two pixels taken at 140 K providing examples of how the algorithm uses the bin size and local maxima to find the RTN peaks within the distribution.}
    \label{fig:linertn}
\end{figure}

Finding RTN-affected pixels involves investigating histograms of CDS values for each individual pixel. If a pixel has one RTN trap, there are two current states and thus three different possible measurement values when performing CDS: a value around zero, a negative value, and a positive value. This means the distinction between a pixel with and without RTN is the number of peaks exhibited in the histogram of its CDS-subtracted pixel values. Each pixel uses a histogram with a pre-selected number of bins that are used for detection based on the read noise of the data being used. Data taken at 140 K have a read noise of 7.9 DN (6.3 e$^{-}$), so bin sizes of 8 DN (6.4 e$^{-}$) are used. Data taken at 160 K have a read noise of 17 DN (14 e$^{-}$), so bin sizes of 16 DN (12.9 e$^{-}$) are used. Each bin center is calculated and the positions of local maxima are determined (see Figure \ref{fig:linertn}). Since there are at least 3 potential measurement values, any pixels that have 2 or fewer local maxima are considered to be likely other types of noise and are removed from further analysis.

Without enough samples, it may be difficult to determine if spurious values are truly RTN or not. A minimum number of pixel values need to be outside of the central distribution for an RTN pattern to become consistent. Simulations were used to test a range of threshold values, where each is a corresponding minimum number of samples that need to be in a single bin. All bins below this threshold are removed from analysis. The threshold value used is presented in the next section. In the case that there are remaining values above this threshold that are not from an RTN state (e.g., two peaks on the right side of the main Gaussian that are the result of an inconsistent pixel), the algorithm determines how the height of each peak is oriented. Since pixels with RTN have peaks with lower magnitudes, distributed symmetrically around the central noise peak, the algorithm finds if this pattern exists.

The algorithm can falsely identify multiple peaks as being a CDS RTN signature due to varying read-noise distributions in a pixel's histogram. A pixel can have a set of exposures that have an increased read noise which might cause those pixel values to appear as a second peak. If a third peak were to form on the opposite side of the central distribution, a triple peak pattern can become present. These patterns can change slowly over time or randomly spike to higher values over time. They do no occur in all pixels, meaning the cause is most likely related to the pixel not performing correctly. We can distinguish these cases from RTN by examining their pattern over time rather than their overall distribution. It has been shown, however, that RTN can appear and disappear over time in certain pixels. In order to find these pixels, we developed a method to detect those specific cases and only remove pixels that have variable read noise states.

This method works by evaluating how the read noise changes between sets of exposures and determining if there is a significant shift in subsequent sets to search for RTN that has appeared. To distinguish small changes in read noise from large changes, we have to determine the statistical significance of the read noise shifts. This is done using bootstrap distributions of the average change in read noise per pixel. A p-value can be found from this distribution, which would determine whether we accept or reject the null hypothesis that states a pixel has a statistically significant change in read noise. We create the bootstrap distribution by taking a sample from the list of changes between read noise values, finding the average, and repeating this 5000 times. A p-value is found by calculating how many times an iteration found a change that was equal to or as extreme as a threshold and dividing that by the total number of iterations. This is used to measure how likely the pixel has inconsistent high read noise states. After some trial-and-error, a threshold of 2 standard deviations was chosen, as this is the point where if enough values are above the threshold, it is most likely  not related to the central Gaussian distribution. A 99.7\% confidence level is pre-determined to be compared to the p-value so we can make a decision on whether or not to accept or reject the null hypothesis (i.e., if the average change in a pixel is greater than 2 standard deviations). If we choose to accept, we can say that the average change in read noise was not significant and the pixel most likely does not have RTN. If we choose to reject, we can say that the change in read noise is significant enough, and we re-evaluate the runs with high read noise for RTN. If it is detected, that pixel is included in the final percentage.

Once a pixel is determined to have RTN signatures, an amplitude and frequency can be found. The amplitude is found by identifying the bin locations of the RTN peaks. Since the precise positions of the peaks within these bins are unknown, we estimate it by drawing a value from a uniform distribution, using boundaries spanned by the bin in which the peak was found. The amplitude of the RTN is taken to be the average of the distance between the RTN peaks and the central peak. The frequency is found by fitting a Gaussian to the central peak and subtracting the number of counts in the fitted Gaussian from the initial histogram. This allows us to remove the central distribution and estimate the number of counts that contribute to the RTN peak. This number can be divided by the total number of frames, which results in the estimated frequency.

The variation in the algorithm is determined by applying a bootstrap sample to each individual pixel's histogram of values. For each sample, the pixel was put through the algorithm to find RTN. This allowed us to provide an estimation of the error of RTN-affected pixels based on how the algorithm detects RTN. These errors are provided along with the final percentages of RTN-affected pixels below. 

\subsubsection{Simulations}

To assess how well the algorithm detects RTN-affected pixels with different characteristics, we create simulated data with unique amplitudes and frequencies of RTN events to assess the sensitivity of our algorithm. First, the number of pixels to simulate is determined based on the size of the detector. Then, a Gaussian distribution is sampled for each pixel using a mean of zero and a unique standard deviation that is randomly sampled from the true read noise distribution of the detector, based on the temperature at which the data were taken. The number of samples from each Gaussian is determined by the number of frames from the true detector data being simulated. A 1024 $\times$ 1024 pixel array with 3,766 frames for data taken at 140 K and 9,214 frames for data taken at 160 K, was generated to determine the false positive rate by simulating data without RTN. The false positive rates at both temperatures are 0.05\% and 0.01\%, respectively. Simulations are unable to be created for the Speedster-EXD550 detectors with a 550 $\times$ 550 pixel array due to memory limitations caused by a significant increase in the number of simulated pixels. Instead, multiple sets of simulated data with smaller array sizes are used with 1,943,000 frames each, and the false positive rate is estimated using these samples. With a 99.7\% confidence level, it can be estimated that the false positive rate for this detector is less than 0.08\%. 

To test how the algorithm detects RTN, each simulation is allotted its own amplitude and frequency for all pixels in the array. Two lists of random numbers are generated with the same number of values as there are frames in each dataset for all pixels. The first list is sampled from a Gaussian distribution with a mean of 0 and the same read noise as the initial dataset. The second list is sampled from a Uniform distribution with values between 0 and 1. We can then generate a dataset with frequency $f$ by finding all values from the first list that correspond with values from the second list that are less than $f$ and flag those points. These pixel values are then given an amplitude that is added 50\% of the time or subtracted 50\% of the time to produce the expected trimodal RTN distribution. The result is a simulated detector with RTN with a given amplitude and frequency in each of the pixels. 

A range of frequency thresholds used in the algorithm is tested for detection accuracy along a range of amplitude and frequency values for each detector. Pixel arrays containing 10,000 pixels are used for the sake of saving time, and the detection rate is noted for each combination. Different temperatures are tested as well and a frequency threshold of 10 frames is chosen as a number that has a good balance between losing the least number of RTN-affected pixels and finding the most number of RTN-affected pixels.

\section{Results}

\subsection{RTN Detection Results}

The Speedster-EXD550 data are processed in the same way as the H2RG data, including the same algorithm parameters for direct comparisons; however, we did not observe the same variation in read noise or RTN that appears and disappears within pixels like we did in the H2RG, so finding the varying read noise states is not necessary for these data. Bootstrap simulations of the Speedster-EXD550 dataset are not feasible due to the large number of frames, so the largest percentage error found on the H2RG data is used. Since the Speedster-EXD550 data have significantly more frames, that error should be less than those found in the H2RG, so the largest percentage error provides a reasonable upper bound. Between both H2RG datasets, the largest error is found to be approximately 3\% of the percentage of RTN affected pixels. For consistency, this percentage error is used for all cases. The final percentages of RTN-affected pixels in the H2RG for both temperatures and both Speedster-EXD550 detectors are presented in Table \ref{tab:rtnpct140}. 

\vspace{0.01cm}

\begin{table}[h]
    \centering
    \begin{tabular}{|l|l|}
    \hline
    Detector & RTN Percentage \\ \hline
    H2RG-122 at 140 K & 0.42 $\pm$ 0.01 \% \\ \hline
    H2RG-122 at 160 K & 0.060 $\pm$ 0.002\% \\ \hline
    Speedster-EXD550 FPM 23056 & 5.1 $\pm$ 0.2 \% \\ \hline
    Speedster-EXD550 FPM 23057 & 7.1 $\pm$ 0.2 \% \\ \hline
    \end{tabular}
    \vspace{0.3cm}
    \caption{Percentage of RTN found in the H2RG at temperatures of 140 K and 160 K and two different Speedster-EXD550 detectors, both taken at 233 K. The error has been approximated using bootstrap methods and represents the highest limitations of the error.}
    \label{tab:rtnpct140}
\end{table}

\begin{figure}[h]
    \centering
    \includegraphics[width=0.87\linewidth]{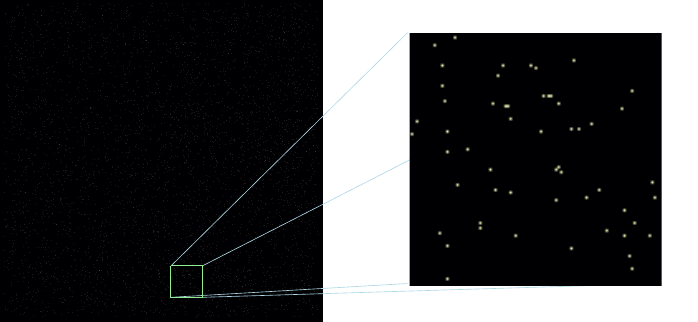}
    \vspace{0.12cm}
    \caption{The spatial distribution of RTN (denoted as yellow dots) in the H2RG at 140 K. The distribution appears to be random, as there are no obvious patterns or clusters of pixels. The right-most plot is a 100 x 100 pixel cutout from the main detector to provide a better idea of the distribution of pixels in the array. This distribution is representative of what is observed in other regions of the detector.}
    \label{fig:det_rtn}
\end{figure}

\begin{figure}[tp]
    \centering
    \includegraphics[width=0.58\linewidth]{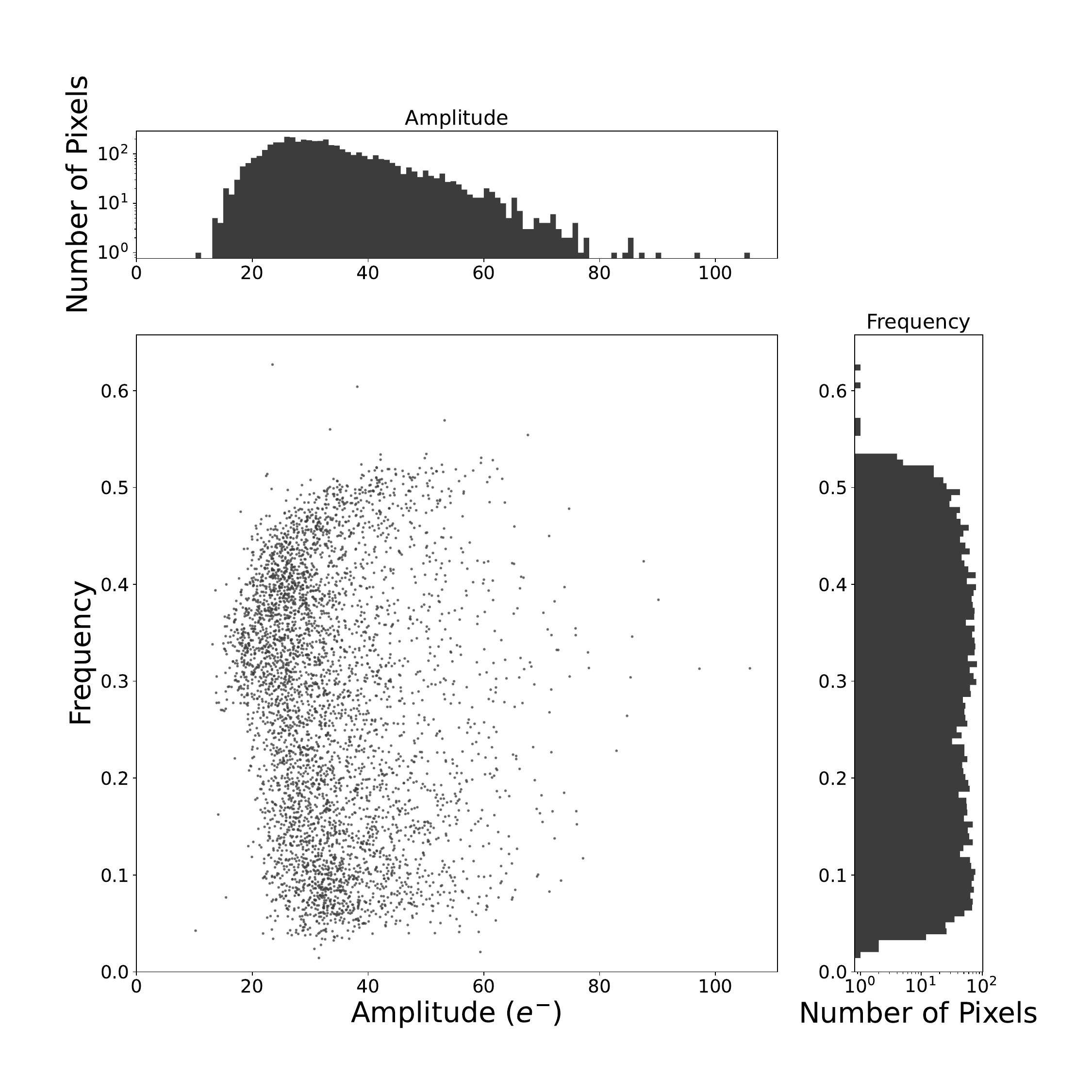}
    \vspace{0.1cm}
    \caption{The distribution of amplitudes and frequencies within the H2RG RTN at 140 K. The same distribution at 160 K included pixels in the same range, but there was no apparent distribution formed due to the lack of pixels.}
    \label{fig:af140}
\end{figure}

\begin{figure}[bp]
    \centering
    \includegraphics[width=0.58\linewidth]{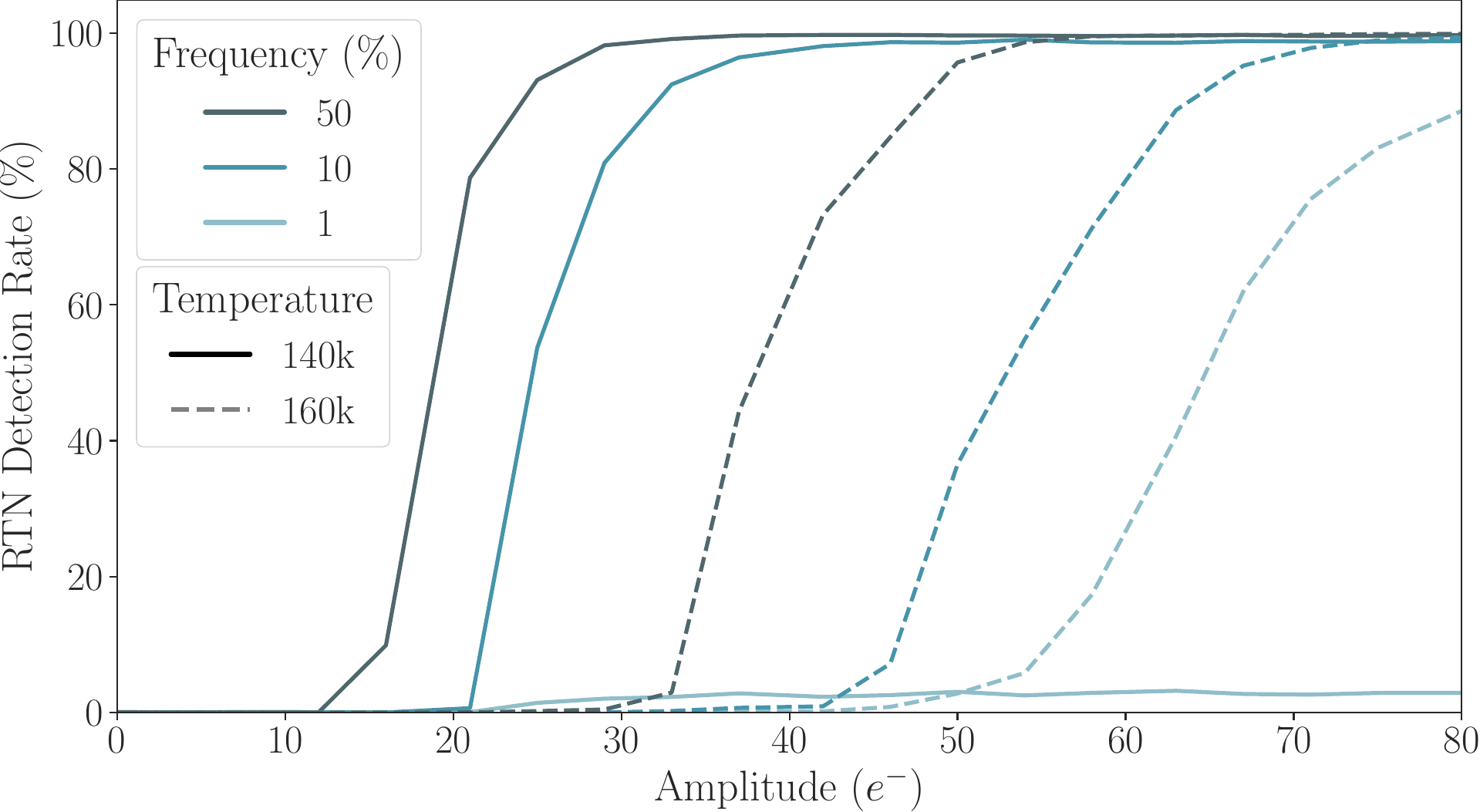}
    \vspace{0.3cm}
    \caption{A plot with the rates of RTN detection over a frequency range and an amplitude range. The solid line indicates the simulated data at 140 K and the dashed line indicates the simulated data at 160 K. It can be seen for both, as the amplitude decreases, the RTN detection rate decreases as well. As the frequency decreases, the amplitude cutoff increases. There is also a significant difference between RTN percentages at 140 K versus 160 K, meaning the read noise has affected these smaller amplitude peaks.}
    \label{fig:detrt}
\end{figure}

\begin{figure}[tp]
    \centering
    \includegraphics[width=0.58\linewidth]{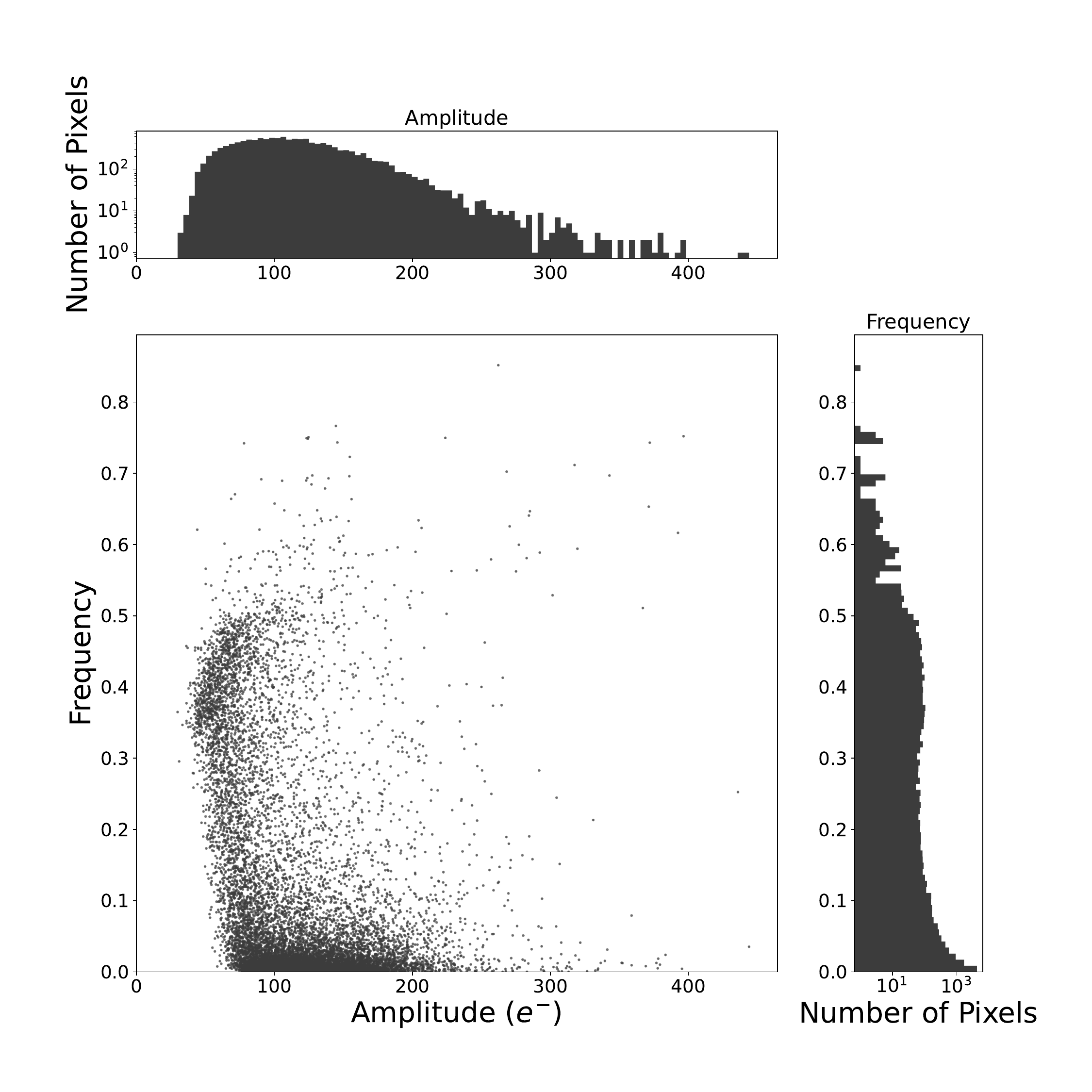}
    \vspace{0.3cm}
    \caption{The distribution of amplitudes and frequencies within the Speedster-EXD550 FPM 23056 detector. The Speedster-EXD550 FPM 23057 detector had the same distribution, but was more dense due to the increase in RTN-affected pixels.}
    \label{fig:afspeed}
\end{figure}

\begin{figure}[bp]
    \centering
    \includegraphics[width=0.5\linewidth]{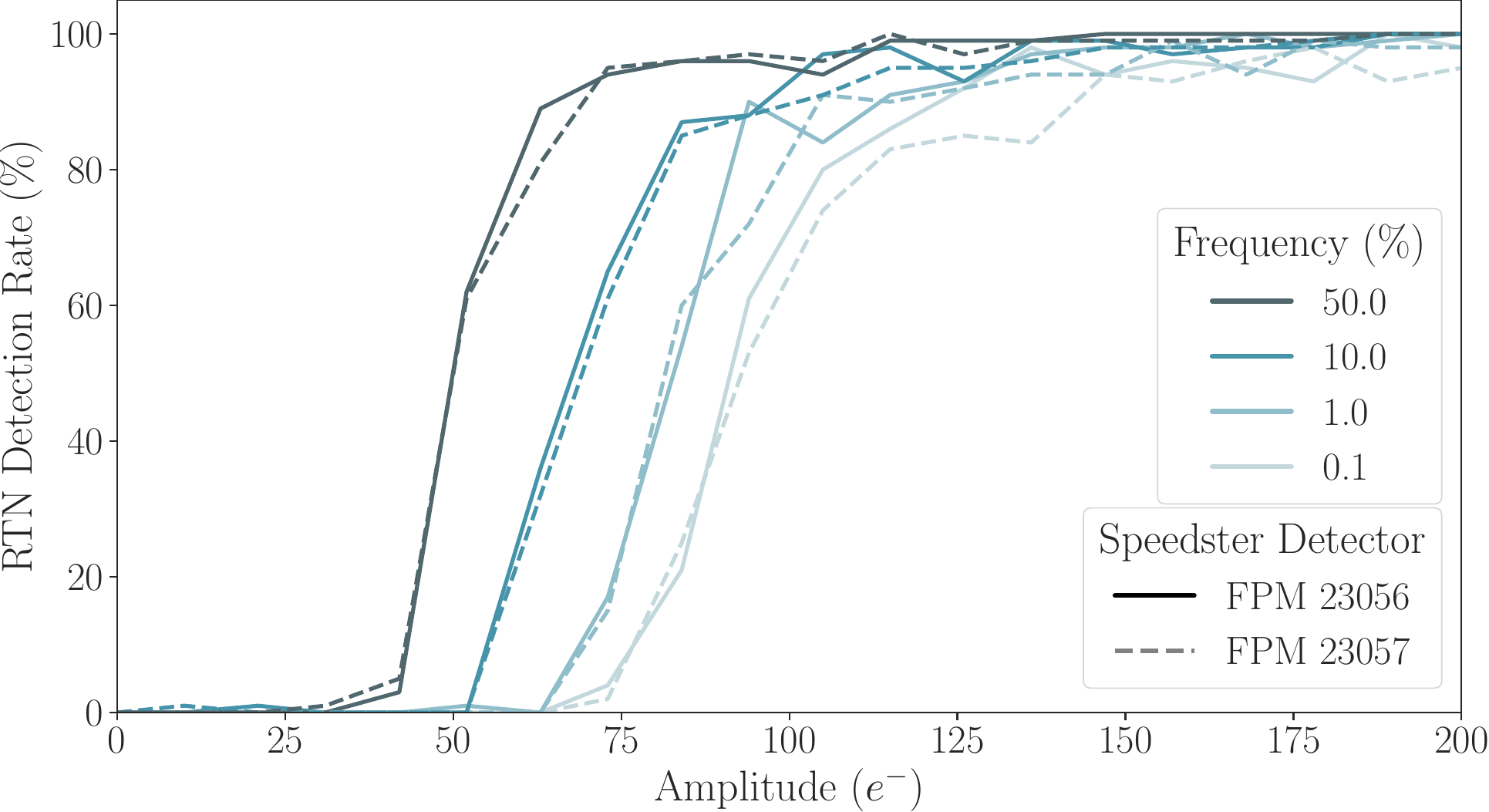}
    \vspace{0.3cm}
    \caption{A plot with the rates of RTN detection over a frequency range and an amplitude range for two Speedster-EXD550 detectors. Just like the H2RG, as the amplitude decreases, the detection rate decreases and as the frequency decreases, the amplitude cutoff increases. With their large amount of frames, these detectors are able to capture much lower frequencies.}
    \label{fig:simspeed}
\end{figure}

The spatial distribution of the RTN in the H2RG at 140 K is shown in Figure \ref{fig:det_rtn}. For the data taken at 160 K, there are not enough points to clearly see the distribution, so this plot is not included. The characteristics of the RTN in this detector are also noted, and a plot of the amplitude and frequency is presented in Figure \ref{fig:af140}. Again, the small amount of data at 160 K did not produce significant results. Simulations are run to test how different amplitudes and frequencies affected the detection rate of the algorithm. Each simulation involved 10,000 pixels at a unique amplitude and frequency. A plot was made to visually assess how the detection rate differs at specific RTN characteristics (see Figure \ref{fig:detrt}). The amplitude and frequency plot for the Speedster-EXD550 detectors is also shown in Figure \ref{fig:afspeed} and simulations were also run for these detectors at different amplitudes and frequencies; the results from these are plotted in Figure \ref{fig:simspeed}. 

\vspace{1cm}

\newpage

\section{Discussion}

\begin{figure}[t]
    \centering
    \includegraphics[width=\linewidth]{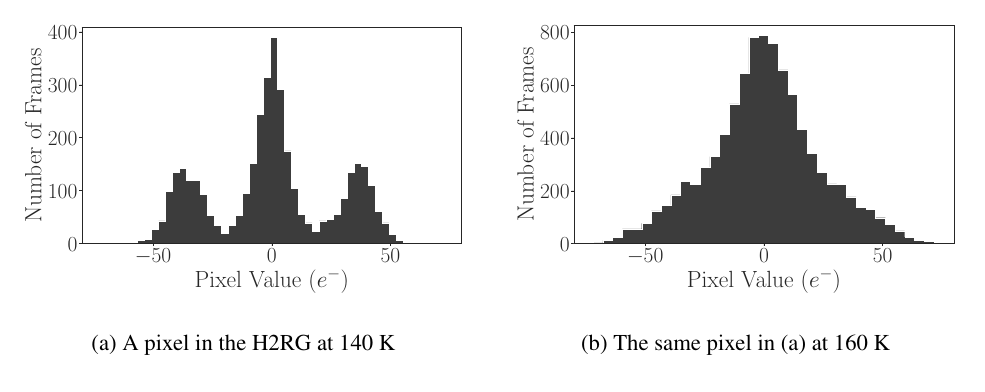}
    \vspace{-0.6cm}
    \caption{These images show the same pixel location in the H2RG at two different temperature states. The increase in temperature caused the read noise to double, which masks the RTN, causing the algorithm to detect only one peak.}
    \label{fig:hiddenrtn}
\end{figure}

From Table \ref{tab:rtnpct140}, there is a significant difference between the percentage of RTN-affected pixels in the H2RG as a function of temperature. This is explained by the increase in read noise at this higher temperature, where a much wider Gaussian is formed that masks low-magnitude RTN within a pixel. Figure \ref{fig:hiddenrtn} shows an example of two identical pixels taken at both temperatures. Note that at the higher temperature, the RTN peaks are blended with the wider central Gaussian. This causes the algorithm to not find any peaks and miss RTN that was obvious at lower temperatures. 

The H2RG data were taken at lower temperatures than the Speedster-EXD550, so we would expect to see a higher amount of RTN in the H2RG if the detectors are affected in the same way. It should also be noted that the Speedster-EXD550 RTN value could potentially be even higher at lower temperatures and/or at lower values of read noise since the higher temperatures and noise could be masking some RTN. Thus, the difference in RTN percentage between H2RG and Speedster-EXD550 would be even more stark. Since we do not see a higher percentage of RTN in the H2RG, it is evident from these RTN observations that the H2RG is impacted by RTN to a lesser degree than than the Speedster-EXD550. These two devices were fabricated at different foundaries, so we speculate that differences in RTN between the two devices may be related to fabrication processes rather than intrinsic design. However, we do not have direct evidence of the cause for RTN differences between the two devices.

The final percentage of RTN-affected pixels consists of only RTN that is detectable given the distribution of the read noise. This is the reason amplitudes below around 12 e$^{-}$ are missing from Figure \ref{fig:af140}. This is also shown in Figure \ref{fig:detrt}, where the simulated RTN have amplitudes around 12 e$^{-}$ that are cutoff, even at frequencies of 50\%. If we wanted to find RTN at low amplitudes, a constant integration time would be needed to see the change in amplifier output over time and conclude its presence. The same effect can be seen in the Speedster-EXD550 plots, where amplitudes below approximately 25 e$^{-}$ are missing and the simulations show the same thing. The amplitude of the RTN in the H2RG appears to be non-Gaussian with a skew and a median amplitude of approximately 30 e$^{-}$. The frequency is almost uniform with ranges between 0\% and slightly above 60\%. The amplitude distribution in the Speedster-EXD550 has much more RTN across a wide range of amplitude space (up to 100's of e$^{-}$) and because of the significant increase in total number of frames, there are also much more low frequency RTN pixels. The frequency distribution in log-space has a unique shape that is different than the H2RG frequency distribution. Although much more overall RTN is found in the Speedster-EXD550 when compared to the H2RG, in particular more lower-frequency RTN and much higher amplitude RTN, the main density of the distribution appears to have some similar gradients across amplitude and frequency space. 

The ability to detect RTN (i.e. ‘detection rate’ or detection efficiency) in the H2RG
(see Figure \ref{fig:detrt}) seems to converge close to 100\% with increasing amplitudes. However, as the amplitudes begin to decrease, the detection rate drops at different points depending on the frequency and temperature. For both higher temperatures and higher frequencies, the amplitude cutoff of detectable RTN is much lower than that of lower temperatures and lower frequencies. It can also be seen that at a frequency of 1\%, the detection rate for data taken at 140 K does not increase with increasing amplitude due to the decrease in amount of frames at these temperatures. The detection rate of RTN in the Speedster-EXD550 (see Figure \ref{fig:simspeed}) also converges close to 100\% with increasing amplitudes. Since there is only one temperature, both detectors are plotted in the same graph and only the frequency needs to be discussed. As seen with the H2RG, as the frequency increases, the amplitude cutoff decreases, revealing more RTN at lower amplitudes. This verifies the claim that more RTN would be found if the amplitude cutoff was below 25 e$^{-}$. With the large number of frames, if we could find RTN below this cutoff, we should find much more RTN in the Speedster-EXD550 detectors. 

Similar methods have been used in the past to determine the amount of RTN in detectors. One experiment on an engineering-grade small-pixel-128 detector found 10-20\% RTN-affected pixels using a very simple RTN detection method. A similar distribution of RTN amplitudes and frequencies was recovered from these data. \cite{hull2021detector} A previous measurement on the same two Speedster-EXD550 detectors used in this analysis, taken at 213 K, found 4.6\% in FPM 23056 and 2.4\% in FPM 23057. \cite{colosimo2023initial}

\section{Conclusion}

These results give confirmation that while the H2RG X-ray HCD does have measurable RTN (0.42\%), it is much lower amplitude and lower frequency than the new Speedster-EXD550 detector ($\sim$5\%). Using archival data from our H2RG X-ray hybrid CMOS detector, we were able to develop methods to characterize the behavior of RTN and its impact on the performance of this detector. Even though it is difficult to detect, it is possible to approximate the percentage of pixels that have high-frequency and high-amplitude that have the greatest impact on detector performance. Peak finding and read noise analysis were the methods used, which have parameters that have been adjusted using simulations. These simulations were made to test the quality of the algorithm and demonstrate its sensitivity to RTN with different characteristics. The first major factor in this uncertainty is the read noise of a detector, which is ultimately affected by the temperature. This effect can be seen by comparing the difference in percentage from data taken at 140 K and 160 K. At higher temperatures, the read noise masks low-magnitude RTN peaks, making them difficult to find. The second factor is the sample size, which impacts the low-frequency detection. This effect can be seen by comparing the difference between low-frequency detection in the H2RG (3,766 frames) and the Speedster-EXD550 (1,943,000 frames). Even though the Speedster-EXD550 data was taken at much higher temperatures (233 K), there is still a reasonable amount of RTN detected.

\subsection*{Disclosures}
No potential conflicts of interest have been identified by the authors.

\subsection* {Code, Data, and Materials Availability} 
Code and data in support of the findings of this paper may be provided upon request. 

\subsection* {Acknowledgments}

We would like to thank Vincent Douence and others at Teledyne Imaging Sensors for helpful conversations and contributions to this work. We greatfully acknowledge support from NASA Grant Nos. 80NSSC18K0147, 80NSSC24K0327, 80NSSC21K1125, and 80NSSC20K0778. 


\bibliography{report}   
\bibliographystyle{spiejour}   


\vspace{2ex}\noindent \textbf{William A. Bevidas Jr.} is a Research Technologist at the Pennsylvania State University in the Department of Astronomy and Astrophyiscs. He received his B.S. degrees of Astronomy \& Astrophysics and Statistics from Penn State University in 2024. His research interests include instrumentation and the application of statistics in astronomy. 

\noindent \textbf{Joseph Colosimo} is a NASA Postdoctoral Program Fellow at NASA Goddard Space Flight Center. He is a member of the Next-Generation X-ray Optics Laboratory, working on the development and characterization of high-resolution X-ray optics. He received his Ph.D. in Astronomy and Astrophysics from the Pennsylvania State University, where he worked on the BlackCAT CubeSat mission and the characterization of X-ray Hybrid CMOS detectors. His research interests include the development and application of technologies to future high-energy astrophysics missions.

\noindent \textbf{Abraham D. Falcone} is an Associate Professor of Astronomy and Astrophysics at Pennsylvania State University as well as the Associate Department Head for the graduate program. He leads research in fields ranging from X-ray and gamma-ray instrumentation to high-energy astrophysics of active galactic nuclei and gamma-ray bursts. He has authored more than 220 refereed publications. He is a member of the Swift team that was awarded the Bruno Rossi Prize in 2007.  He is the Principal Investigator of the NASA BlackCAT CubeSat. He is also a VERITAS collaboration member and a collaborator on multiple future X-ray and gamma-ray astronomy mission development efforts including AXIS, Athena, MoonCAT, and Lynx.

\vspace{1ex}
\noindent Biographies of the other authors are not available.

\listoftables
\listoffigures

\end{spacing}
\end{document}